\documentclass{llncs}

\usepackage{amssymb,stmaryrd}

\def\vde{\vdash_{\equiv}}

\newbox\tempa
\newbox\tempb
\newdimen\tempc
\def\mud#1{\hfil $\displaystyle{\mathstrut #1}$\hfil}
\def\rig#1{\hfil $\displaystyle{#1}$}
\def\irulehelp#1#2#3{\setbox\tempa=\hbox{$\displaystyle{\mathstrut #2}$}%
                        \setbox\tempb=\vbox{\halign{##\cr
        \mud{#1}\cr
        \noalign{\vskip\the\lineskip}
        \noalign{\hrule height 0pt}
        \rig{\vbox to 0pt{\vss\hbox to 0pt{${\; #3}$\hss}\vss}}\cr
        \noalign{\hrule}
        \noalign{\vskip\the\lineskip}
        \mud{\copy\tempa}\cr}}
                      \tempc=\wd\tempb
                      \advance\tempc by \wd\tempa
                      \divide\tempc by 2 }
\def\irule#1#2#3{{\irulehelp{#1}{#2}{#3}
                     \hbox to \wd\tempa{\hss \box\tempb \hss}}}

\begin{document}

\title{The Stratified Foundations as a theory modulo}

\author{Gilles Dowek}

\institute{INRIA-Rocquencourt\\
B.P. 105, 78153 Le Chesnay Cedex, France. \\
{\tt Gilles.Dowek@inria.fr}, {\tt http://logical.inria.fr/\~{}dowek}}

\date{}

\maketitle

\begin{abstract}
The {\em Stratified Foundations} are a restriction of naive set theory
where the comprehension scheme is restricted to stratifiable
propositions.  It is known that this theory is consistent and that
proofs strongly normalize in this theory.  {\em Deduction modulo} is a
formulation of first-order logic with a general notion of cut. It is
known that proofs normalize in a theory modulo if it has some kind of
many-valued model called a {\em pre-model}.  We show in this paper
that the Stratified Foundations can be presented in deduction modulo
and that the method used in the original normalization proof can be
adapted to construct a pre-model for this theory.
\end{abstract}

The {\em Stratified Foundations} are a restriction of naive set theory
where the comprehension scheme is restricted to stratifiable
propositions. This theory is consistent \cite{Jensen} while naive set theory 
is not and the consistency of the Stratified Foundations
together with the extensionality axiom -~the so-called {\em New
Foundations}~- is open.

The Stratified Foundations extend simple type theory and, like in simple
type theory, proofs strongly normalize in The Stratified Foundations
\cite{Crabbe91}.
These two normalization 
proofs, like many, have some
parts in common, for instance they both use Girard's reducibility candidates. 
This motivates the investigation of general
normalization theorems that have normalization theorems for specific
theories as consequences.  The normalization theorem for deduction
modulo \cite{DowekWerner-normalisation-98} is an example of such a
general theorem. It concerns theories expressed in {\em deduction
modulo} \cite{DHK-ENAR-98} that are first-order theories with a
general notion of cut. According to this theorem, proofs normalize in
a theory in deduction modulo if this theory has some kind of
many-valued model called a {\em pre-model}.  For instance, simple type
theory can be expressed in deduction modulo
\cite{DHK-ENAR-98,DHK-HOL-ls-rta99} and it has a pre-model
\cite{DowekWerner-normalisation-98,DHK-HOL-ls-rta99} and hence it has
the normalization property.  The normalization proof obtained this way
is modular: all the lemmas specific to type theory are concentrated in
the pre-model construction while the theorem that the existence of a
pre-model implies normalization is generic and can be used for any
other theory in deduction modulo.

The goal of this paper is to show that the Stratified Foundations also
can be presented in deduction modulo and that the method used in
the original normalization proof can be adapted to construct a
pre-model for this theory.  The normalization proof obtained this
way is simpler than the original one because it simply uses the fact
that proofs normalize in the Stratified Foundations 
if this theory has a pre-model, while a variant of this proposition needs to be
proved in the original proof.

It is worth noticing that the original normalization proof for the
Stratified Foundations is already in two steps, where the first is the
construction of a so-called {\em normalization model} and the second
is a proof that proofs normalize in the Stratified Foundations 
if there is such a normalization model. Normalization models are, 
more or less, pre-models of the Stratified Foundations.
So, we show that the notion of normalization model,
that is specific to the Stratified Foundations, is an instance of a
more general notion that can be defined 
for all theories modulo, and that the lemma that the existence of a
normalization model implies normalization for the Stratified
Foundations is an instance of a more general theorem that holds for
all theories modulo. 

The normalization proof obtained this way differs also from the
original one in other respects. First, to remain in first-order logic,
we do not use a presentation of the Stratified Foundations with a
binder, but one with combinators. To express the Stratified
Foundations with a binder in first-order logic, we could use de Bruijn
indices and explicit substitutions along the lines of
\cite{DHK-HOL-ls-rta99}. The pre-model construction below should
generalize easily to such a presentation.  Second, our cuts are cuts
modulo, while the original proof uses Prawitz' {\em folding-unfolding}
cuts.  It is shown in \cite{Dowekcoupures} that the normalization
theorems are equivalent for the two notions of cuts, but that the
notion of cut modulo is more general that the notion of
folding-unfolding cut. Third, we use untyped reducibility candidates 
and not typed ones as in the original proof. This quite simplifies the
technical details.

A last benefit of expressing the Stratified Foundations in deduction
modulo is that we can use the method developed in \cite{DHK-ENAR-98} to 
organize proof search. The method obtained this way, that is an analog
of higher-order resolution for the Stratified Foundations, is much more
efficient than usual first-order proof search methods with the
comprehension axioms, although it remains complete as the Stratified
Foundations have the normalization property. 

\section{Deduction modulo}

\subsection{Identifying propositions}

In deduction modulo, the notions of language, term and proposition are
that of first-order logic. But, a theory is formed with a set of
axioms $\Gamma$ {\em and a congruence $\equiv$} defined on
propositions. Such a congruence may be defined by a rewrite systems on
terms and on propositions (as propositions contain binders -~quantifiers~-, 
these rewrite systems are in fact {\em combinatory
reduction systems} \cite{KlopOostromRaamsdonk}).  Then, the deduction
rules take this congruence into account. For instance, the {\em modus
ponens} is not stated as usual
$$\irule{A \Rightarrow B~~~A}{B}{}$$
as the first premise need not be exactly $A \Rightarrow B$ but may be
only congruent to this proposition, hence it is stated
$$\irule{C~~~A}{B}{\mbox{if $C \equiv A \Rightarrow B$}}$$

\begin{figure}
\noindent

{
\hspace*{3cm}
$\irule{}
        {\Gamma \vde B}
        {\mbox{axiom if $A \in \Gamma$ and $A \equiv B$}}$

\smallskip 
\hspace*{3cm}
$\irule{\Gamma, A \vde B}
        {\Gamma \vde C}
        {\mbox{$\Rightarrow$-intro if $C \equiv (A
        \Rightarrow B)$}}$

\smallskip 
\hspace*{3cm}
$\irule{\Gamma \vde C~~~\Gamma \vde A}
        {\Gamma \vde B}
        {\mbox{$\Rightarrow$-elim if $C \equiv (A \Rightarrow B)$}}$
 
\smallskip 
\hspace*{3cm}
$\irule{\Gamma \vde A~~~\Gamma \vde B}
        {\Gamma \vde C}
        {\mbox{$\wedge$-intro if $C \equiv (A \wedge B)$}}$

\smallskip 
\hspace*{3cm}
$\irule{\Gamma \vde C}
        {\Gamma \vde A}
        {\mbox{$\wedge$-elim if $C \equiv (A \wedge B)$}}$
 
\smallskip 
\hspace*{3cm}
$\irule{\Gamma \vde C}
        {\Gamma \vde B}
        {\mbox{$\wedge$-elim if $C \equiv (A \wedge B)$}}$

\smallskip 
\hspace*{3cm}
$\irule{\Gamma \vde A}
        {\Gamma \vde C}
        {\mbox{$\vee$-intro if $C \equiv (A \vee B)$}}$
 
\smallskip 
\hspace*{3cm}
$\irule{\Gamma \vde B}
        {\Gamma \vde C}
        {\mbox{$\vee$-intro if $C \equiv (A \vee B)$}}$
 
\smallskip 
\hspace*{3cm}
$\irule{\Gamma \vde D~~~\Gamma, A \vde C~~~\Gamma, B \vde C}
        {\Gamma \vde C}
        {\mbox{$\vee$-elim if $D \equiv (A \vee B)$}}$
 
\smallskip 
\hspace*{3cm}
$\irule{\Gamma \vde B}
        {\Gamma \vde A}
        {\mbox{$\bot$-elim if $B \equiv \bot$}}$
 
\smallskip 
\hspace*{3cm}
$\irule{\Gamma \vde A}
        {\Gamma \vde B}
        {\mbox{$(x,A)$ $\forall$-intro if $B \equiv (\forall x~A)$ and
                         $x \not\in FV(\Gamma)$}}$
 
\smallskip 
\hspace*{3cm}
$\irule{\Gamma \vde B}
        {\Gamma \vde C}
        {\mbox{$(x,A,t)$ $\forall$-elim if $B \equiv (\forall x~A)$ and $C
         \equiv [t/x]A$}}$
 
\smallskip 
\hspace*{3cm}
$\irule{\Gamma \vde C}
        {\Gamma \vde B}
        {\mbox{$(x,A,t)$ $\exists$-intro if $B \equiv (\exists x~A)$ and $C
               \equiv [t/x]A$}}$
 
\smallskip 
\hspace*{3cm}
$\irule{\Gamma \vde C~~~\Gamma, A \vde B}
        {\Gamma \vde B}
        {\mbox{$(x,A)$ $\exists$-elim if $C \equiv (\exists x~A)$ and
               $x \not\in FV(\Gamma B)$}}$

\smallskip 
\hspace*{3cm}
$\irule{}
        {\Gamma \vde A}
        {\mbox{$B$~Excluded middle if $A \equiv B \vee (B \Rightarrow \bot)$}}$

\caption{Natural deduction modulo}
\label{NatMod}
}

\end{figure}

All the rules of intuitionistic natural deduction may be stated in a
similar way. Classical deduction modulo is obtained by adding the
{\em excluded middle} rule (see figure \ref{NatMod}).

For example, in arithmetic, we can define a congruence with the
following rewrite system 
$$0 + y \rightarrow y$$
$$S(x) + y \rightarrow S(x+y)$$
$$0 \times y \rightarrow 0$$
$$S(x) \times y \rightarrow x \times y + y$$
In the theory formed with a set of axioms $\Gamma$ containing the axiom
$\forall x~x = x$ and this congruence, we can prove, in natural deduction
modulo, that the 
number $4$ is even
$$\irule{\irule{\irule{}
                      {\Gamma \vde \forall x~x = x}
                      {\mbox{axiom}}
               }
               {\Gamma \vde 2 \times 2 = 4}
               {\mbox{$(x,x = x,4)$ $\forall$-elim}}
        }
        {\Gamma \vde  \exists x~2 \times x = 4}
        {\mbox{$(x,2 \times x = 4,2)$ $\exists$-intro}}$$
Substituting the variable $x$ by the term $2$ in the proposition 
$2 \times x = 4$ yields the proposition $2 \times 2 = 4$, 
that is congruent to $4 = 4$. The transformation of one proposition
into the other, that requires several proof steps in usual natural
deduction, is dropped from the proof in deduction modulo.

In this example, all the rewrite rules apply to terms. Deduction
modulo permits also to consider rules rewriting atomic
propositions to arbitrary ones. For instance, in the theory of
integral domains, we have the rule 
$$x \times y = 0 \rightarrow x = 0 \vee y = 0$$
that rewrites an atomic proposition to a disjunction.

Notice that, in the proof above, we do not need the axioms of addition 
and multiplication. Indeed, these axioms are now redundant: since the
terms $0 + y$ and $y$ are congruent, the axiom $\forall y~0 + y = y$ is
congruent to the axiom of equality $\forall y~y = y$. Hence, it can be
dropped. Thus, rewrite rules replace axioms. 

This equivalence between rewrite rules and axioms is expressed by the
the {\em equivalence lemma} that for every congruence $\equiv$, we can
find a theory ${\cal T}$ such that $\Gamma \vde A$ is provable in
deduction modulo if and only if ${\cal T} \Gamma \vdash A$ is provable
in ordinary first-order logic \cite{DHK-ENAR-98}. Hence, deduction
modulo is not a true extension of first-order logic, but rather an
alternative formulation of first-order logic. Of course, the provable
propositions are the same in both cases, but the proofs are very
different.

\subsection{Model of a theory modulo}

A {\em model} of a congruence $\equiv$ is a model such that if $A
\equiv B$ then for all assignments, $A$ and $B$ have the same
denotation.  A {\em model} of a theory modulo $\Gamma, \equiv$ is a
model of the theory $\Gamma$ and of the congruence $\equiv$.
Unsurprisingly, the completeness theorem extends to classical
deduction modulo \cite{Dowek-habilitation} and a proposition is
provable in the theory $\Gamma, \equiv$ if and only if it is valid in
all the models of $\Gamma, \equiv$.

\subsection{Normalization in deduction modulo}

Replacing axioms by rewrite rules in a theory changes the structure of
proofs and in particular some theories may have the normalization
property when expressed with axioms and not when expressed with
rewrite rules.
For instance, from the normalization theorem for first-order logic, 
we get that any proposition that is provable with the axiom $A
\Leftrightarrow (B \wedge (A \Rightarrow \bot))$ has a normal proof.
But if we transform this axiom into the rule 
$A \rightarrow B \wedge  (A \Rightarrow \bot)$
(Crabb\'e's rule \cite{Crabbe74}) the proposition $B \Rightarrow \bot$ has a
proof, but no normal proof. 

We have proved a {\em normalization theorem}: proofs normalize in a
theory modulo if this theory has a {\em pre-model}
\cite{DowekWerner-normalisation-98}. A pre-model is a many-valued
model whose truth values are reducibility candidates, i.e. sets of
proof-terms. Hence we first define proof-terms, then reducibility
candidates and at last pre-models.

\begin{definition}[Proof-term]

{\em Proof-terms} are inductively defined as follows.
\begin{tabbing}
$\pi ::=$ \= $~~\alpha$\\
          \> $|~ \lambda \alpha~\pi ~|~ (\pi~\pi')$\\
          \> $|~ \langle \pi,\pi' \rangle ~|~ fst(\pi) ~|~ snd(\pi)$\\
          \> $|~ i(\pi) ~|~ j(\pi) ~|~ (\delta~\pi_{1}~\alpha \pi_{2}~\beta \pi_{3})$\\
          \> $|~ (botelim~\pi)$\\
          \> $|~ \lambda x~\pi ~|~ (\pi~t)$\\
          \> $|~ \langle t,\pi \rangle~|~ (exelim~\pi~x \alpha \pi')$
\end{tabbing}
\end{definition}

Each proof-term construction corresponds to an intuitionistic 
natural deduction rule:
terms of the form $\alpha$ express proofs built with the axiom rule, 
terms of the form $\lambda \alpha~\pi$ and $(\pi~\pi')$ express proofs
built with the introduction and elimination rules of the implication, 
terms of the form $\langle \pi,\pi' \rangle$ and $fst(\pi)$, $snd(\pi)$ express
proofs built with the introduction and elimination rules of the
conjunction, terms of the form $i(\pi), j(\pi)$ and
$(\delta~\pi_{1}~\alpha \pi_{2}~\beta \pi_{3})$ express proofs built
with the introduction and elimination rules of the disjunction, terms of
the form $(botelim~\pi)$ express proofs built with the elimination
rule of the contradiction, terms of the form 
$\lambda x~\pi$ and $(\pi~t)$ express proofs built with the
introduction and elimination rules of the universal quantifier and terms of
the form $\langle t,\pi \rangle$ and $(exelim~\pi~x \alpha \pi')$ express proofs
built with the introduction and elimination rules of the existential
quantifier. 

\begin{definition}[Reduction]
{\em Reduction} on proof-terms is defined by the following
rules that eliminate cuts step by step.
$$(\lambda \alpha~\pi_{1}~\pi_{2}) \triangleright [\pi_{2}/\alpha]\pi_{1}$$
$$fst(\langle \pi_{1},\pi_{2} \rangle) \triangleright \pi_{1}$$
$$snd(\langle \pi_{1},\pi_{2} \rangle) \triangleright \pi_{2}$$
$$(\delta~i(\pi_{1})~\alpha \pi_{2}~\beta \pi_{3}) 
\triangleright [\pi_{1}/\alpha]\pi_{2}$$
$$(\delta~j(\pi_{1})~\alpha \pi_{2}~\beta \pi_{3}) 
\triangleright [\pi_{1}/\beta]\pi_{3}$$
$$(\lambda x~\pi~t) \triangleright [t/x]\pi$$
$$(exelim~\langle t,\pi_{1} \rangle~\alpha x\pi_{2}) \triangleright
[t/x,\pi_{1}/\alpha]\pi_{2}$$
\end{definition}

\begin{definition}[Reducibility candidates]
A proof-term is said to be {\em neutral} 
if it is a proof variable or an elimination (i.e. of the form 
$(\pi~\pi')$, $fst(\pi)$, $snd(\pi)$, $(\delta~\pi_{1}~\alpha
\pi_{2}~\beta \pi_{3})$, $(botelim~\pi)$, $(\pi~t)$, 
$(exelim~\pi~x \alpha \pi')$), but not an introduction. A set
$R$ of proof-terms is a {\em reducibility candidate} if 
\begin{itemize}
\item if $\pi \in R$, then $\pi$ is strongly normalizable,
\item if $\pi \in R$ and $\pi \triangleright \pi'$ then $\pi' \in R$,
\item if $\pi$ is neutral and 
if for every $\pi'$ such that $\pi \triangleright^{1} \pi'$, $\pi' \in R$ then 
$\pi \in R$. 
\end{itemize}
\end{definition}

We write ${\cal C}$ for the set of all reducibility candidates.

\begin{definition}[Pre-model]
A {\em pre-model} ${\cal N}$ for a language ${\cal L}$ is given by: 
\begin{itemize}
\item a set $N$, 
\item for each function symbol $f$ of arity $n$ a function $\hat{f}$ 
from $N^{n}$ to $N$, 
\item for each predicate symbol $P$ a
function $\hat{P}$ 
from $N^{n}$ to ${\cal C}$.
\end{itemize}
\end{definition}

\begin{definition}[Denotation in a pre-model]
Let ${\cal N}$ be a pre-model, $t$ be a term and $\varphi$ an
assignment mapping all the 
free variables of $t$ to elements of $N$. 
We define the object $\llbracket t \rrbracket^{\cal N}_{\varphi}$ by induction over the
structure of $t$.
\begin{itemize}
\item $\llbracket x\rrbracket^{\cal N}_{\varphi} = \varphi(x)$, 
\item $\llbracket f(t_{1}, \ldots, t_{n})\rrbracket^{\cal N}_{\varphi} = \hat{f}(\llbracket t_{1}\rrbracket^{\cal N}_{\varphi},
\ldots, \llbracket t_{n}\rrbracket^{\cal N}_{\varphi})$.
\end{itemize}

Let $A$ be a proposition and $\varphi$ an assignment mapping all the
free variables of $A$ to elements of $N$. 
We define the reducibility candidate $\llbracket
A\rrbracket^{\cal N}_{\varphi}$ by induction over the structure of $A$.

\begin{itemize}
\item
If $A$ is an atomic proposition $P(t_{1}, \ldots, t_{n})$ then 
$\llbracket A\rrbracket^{\cal N}_{\varphi} = 
\hat{P}(\llbracket t_{1}\rrbracket^{\cal N}_{\varphi}, \ldots, \llbracket
t_{n}\rrbracket^{\cal N}_{\varphi})$.

\item
If $A = B \Rightarrow C$ then 
$\llbracket A\rrbracket^{\cal N}_{\varphi}$ is the set of proofs $\pi$ such
that $\pi$ is strongly normalizable and whenever it
reduces to $\lambda \alpha~\pi_{1}$ then for every
$\pi'$ in $\llbracket B\rrbracket^{\cal N}_{\varphi}$, $[\pi' /
\alpha]\pi_{1}$ is in $\llbracket C\rrbracket^{\cal N}_{\varphi}$.

\item
If $A = B \wedge C$ then 
$\llbracket A\rrbracket^{\cal N}_{\varphi}$ is the set of proofs $\pi$ such
that $\pi$ is strongly normalizable and whenever
it reduces to $\langle \pi_{1},\pi_{2} \rangle$ then
$\pi_{1}$ is in $\llbracket B\rrbracket^{\cal N}_{\varphi}$ and 
$\pi_{2}$ is in $\llbracket C\rrbracket^{\cal N}_{\varphi}$.

\item
If $A = B \vee C$ then 
$\llbracket A\rrbracket^{\cal N}_{\varphi}$ is the set of proofs $\pi$ such
that $\pi$ is strongly normalizable and whenever
it reduces to $i(\pi_{1})$ (resp. $j(\pi_{2})$) then
$\pi_{1}$ (resp. $\pi_{2}$) is in $\llbracket
B\rrbracket^{\cal N}_{\varphi}$ (resp. $\llbracket C\rrbracket^{\cal N}_{\varphi}$).

\item
If $A = \bot$ then 
$\llbracket A\rrbracket^{\cal N}_{\varphi}$ is the set of strongly normalizable
proofs.

\item
If $A = \forall x~B$ then 
$\llbracket A\rrbracket^{\cal N}_{\varphi}$ is the set of proofs $\pi$ such
that $\pi$ is strongly normalizable and whenever it
reduces to $\lambda x~\pi_{1}$ then for every term
$t$ and every element $a$ of $N$ $[t/x]\pi_{1}$ is in
$\llbracket B\rrbracket^{\cal N}_{\varphi+ a/x}$.

\item
If $A = \exists x~B$ then 
$\llbracket A\rrbracket^{\cal N}_{\varphi}$ is the set of proofs $\pi$ such
that $\pi$ is strongly normalizable and whenever
it reduces to $\langle t,\pi_{1} \rangle$ then there exists an element
$a$ in $N$ such that $\pi_{1}$ is in $\llbracket
B\rrbracket^{\cal N}_{\varphi + a/x}$.
\end{itemize}
\end{definition}

\begin{definition}
A pre-model is said to be a {\em pre-model of a congruence $\equiv$}
if when $A \equiv B$ then for 
every assignment $\varphi$, $\llbracket A\rrbracket^{\cal N}_{\varphi} = \llbracket B\rrbracket^{\cal N}_{\varphi}$.  
\end{definition}

\begin{theorem}[Normalization] 
\cite{DowekWerner-normalisation-98}
If a congruence $\equiv$ has a pre-model 
all proofs modulo $\equiv$ strongly normalize. 
\end{theorem}

\section{The Stratified Foundations}

\subsection{The Stratified Foundations as a first-order theory}

\begin{definition} (Stratifiable proposition)

A proposition $A$ in the language $\in$ is said to be {\em
stratifiable} if there exists a 
function $S$ mapping every variable (bound or free) of $A$ to a
natural number in such a way that every atomic proposition of $A$, $x
\in y$ is such that $S(y) = S(x) + 1$. 
\end{definition}

For instance, the proposition 
$$\forall v~(v \in x \Leftrightarrow v \in y) \Rightarrow 
\forall w~(x \in w \Rightarrow y \in w)$$
is stratifiable (take, for instance, $S(v) = 4$, $S(x) = S(y) = 5$,
$S(w) = 6$) but not the proposition 
$$\forall v~(v \in x \Leftrightarrow v \in y) \Rightarrow x \in y$$

\begin{definition} (The stratified comprehension scheme)

For every stratifiable proposition $A$ whose free variables are among
$x_{1}, \ldots, x_{n}, x_{n+1}$ we take the axiom 
$$\forall x_{1}~\ldots~\forall x_{n}~\exists z~\forall x_{n+1}~(x_{n+1} \in z
\Leftrightarrow A)$$ 
\end{definition}

\begin{definition} (The skolemized stratified comprehension scheme)

When we skolemize this scheme, we introduce for each stratifiable
proposition $A$ in the language $\in$ and sequence of variables
$x_{1}, \ldots, x_{n}, x_{n+1}$ such that the free variables of $A$ are among 
$x_{1}, \ldots, x_{n}, x_{n+1}$, a function symbol $f_{x_{1}, \ldots, x_{n},
x_{n+1}, A}$ and the axiom 
$$\forall x_{1}~\ldots~\forall x_{n}~\forall x_{n+1}~(x_{n+1} \in f_{x_{1}, \ldots, 
x_{n}, x_{n+1}, A}(x_{1}, \ldots, x_{n}) \Leftrightarrow A)$$ 
\end{definition}

\subsection{The Stratified Foundations as a theory modulo}

Now we want to replace the axiom scheme above by a rewrite rule,
defining a congruence on propositions, so that the Stratified Foundations
are defined as an axiom free theory modulo. 

\begin{definition} (The rewrite system ${\cal R}$)
$$t_{n+1} \in f_{x_{1}, \ldots, x_{n}, x_{n+1}, A}(t_{1}, \ldots, t_{n}) 
\rightarrow [t_{1}/x_{1}, \ldots, t_{n}/x_{n},t_{n+1}/x_{n+1}]A$$
\end{definition}

\begin{proposition}
The rewrite system ${\cal R}$ is confluent and terminating.
\end{proposition}

\proof{The system ${\cal R}$ is an orthogonal combinatory reduction system,
hence it is confluent \cite{KlopOostromRaamsdonk}. 

For termination, if $A$ is an atomic proposition we write $\|A\|$ for the 
number of function symbols in $A$ and if $A$ is a proposition containing
the atomic propositions $A_{1}, \ldots, A_{p}$ we write
$A^{\circ}$ for the multiset $\{\|A_{1}\|, \ldots,
\|A_{p}\|\}$. We show that if a proposition $A$ reduces in one step to
a proposition $B$ then $B^{\circ} < A^{\circ}$ for the multiset
ordering.

If the proposition $A$ reduces in one step to $B$, 
there is an atomic proposition of $A$, say $A_{1}$, that has the form 
$t_{n+1} \in f_{x_{1}, \ldots, x_{n}, x_{n+1}, C}(t_{1}, \ldots, t_{n})$ and
reduces to $B_{1} = [t_{1}/x_{1}, \ldots, t_{n}/x_{n}, t_{n+1}/x_{n+1}]C$.
Every atomic proposition $b$ of $B_{1}$ has the form
$[t_{1}/x_{1}, \ldots, t_{n}/x_{n}, t_{n+1}/x_{n+1}]c$  
where $c$ is an atomic proposition of $C$. 
The proposition $c$ has the
form $x_{i} \in x_{j}$ for distinct $i$ and $j$ (since $C$ is
stratifiable) $x_{i} \in y$, $y \in x_{i}$ or $y \in z$. 
Hence  
$b$ has the form $t_{i} \in t_{j}$ for distinct $i$ and $j$, 
$t_{i} \in y$, $y \in t_{i}$ or $y \in z$ and $\|b\| < \|A_{1}\|$.
Therefore $B^{\circ} < A^{\circ}$.}

\begin{proposition}
A proposition $A$ is provable from the skolemized comprehension scheme
if and only if it is provable modulo the rewrite system ${\cal R}$. 
\end{proposition}

\subsection{Consistency}

We want now to construct a model for the Stratified Foundations.

If ${\cal M}$ is a model of set theory we write $M$ for the set of elements
of the model, $\in_{\cal M}$ for the denotation of the symbol $\in$ in
this model, $\wp_{\cal M}$ for the powerset in this model, etc.
We write also $\llbracket A \rrbracket^{\cal M}_{\varphi}$ for the denotation of a
proposition $A$ for the assignment $\varphi$. 

The proof of the consistency of the Stratified Foundations rests on
the existence of a model of Zermelo's set theory, such that there is a
bijection $\sigma$ from $M$ to $M$ and a family $v_{i}$ of elements
of $M$, $i \in {\mathbb Z}$ such that
$$a \in_{\cal M} b~\mbox{if and only if}~\sigma a \in_{\cal M} \sigma b$$
$$\sigma v_{i} = v_{i+1}$$ 
$$v_{i} \subseteq_{\cal M} v_{i+1}$$
$$\wp_{\cal M} (v_{i}) \subseteq_{\cal M} v_{i+1}$$
The existence of such a model is proved in \cite{Jensen}. 

Using the fact that ${\cal M}$ is a model of the axiom of
extensionality, we prove that 
$a \subseteq_{\cal M} b$ if and only if 
$\sigma a \subseteq_{\cal M} \sigma b$, 
$\sigma \{a,b\}_{\cal M} = 
\{\sigma a, \sigma b\}_{\cal M}$, 
$\sigma \langle a,b \rangle_{\cal M} = 
\langle \sigma a, \sigma b \rangle_{\cal M}$, 
$\sigma \wp(a) = \wp (\sigma a)$, etc.

For the normalization proof, we will further need that ${\cal M}$
is an $\omega$-model.   
We define $\overline{0} = \emptyset_{\cal M}$, 
$\overline{n+1} = \overline{n} \cup_{\cal M}
\{\overline{n}\}_{\cal M}$. 
An $\omega$-model is a model such that 
$a \in_{\cal M} {\mathbb N}_{\cal M}$ if and only if there exists $n$
in ${\mathbb N}$ such that $a = \overline{n}$. 
The existence of such a model is proved in \cite{Jensen} (see
also \cite{Crabbe91}). 

Using the fact that ${\cal M}$ is a model of the axiom of
extensionality, we prove that $\sigma \emptyset_{\cal M} =
\emptyset_{\cal M}$ and then, by induction on $n$ that 
$\sigma \overline{n} = \overline{n}$. 

Notice that since $\wp_{\cal M} (v_{i}) \subseteq_{\cal M} v_{i+1}$, 
$\emptyset_{\cal M} \in_{\cal M} v_{i}$ and for all $n$, $\overline{n}
\in_{\cal M} v_{i}$. Hence as the model is an $\omega$-model ${\mathbb
N}_{\cal M} \subseteq_{\cal M} v_{i}$. 

In an $\omega$-model, we can identify the set ${\mathbb N}$ of natural
numbers with the set of objects $a$ in ${\cal M}$ such that $a
\in_{\cal M} {\mathbb N}_{\cal M}$. To each proof-term 
we can associate a natural number $n$ (its
G\"odel number) and then the element $\overline{n}$ of ${\cal
M}$. Proof-terms, their G\"odel number and the encoding of this number
in ${\cal M}$ will be identified in the following.

\medskip

We are now ready to construct a model ${\cal U}$ for the Stratified
Foundations. The base set is the set $U$ of elements $a$ of $M$ such
that $a \in_{\cal M} v_{0}$.
The relation $\in_{\cal U}$ is defined by $a~\in_{\cal U}~b$ if and
only if $a \in_{\cal M} \sigma b$.  This permits to define the
denotation of propositions built without Skolem symbols.  
To be able to define the denotation of Skolem symbols, we
prove the following proposition.

\begin{proposition}
For every stratifiable proposition $A$ in the language $\in$
whose free variables are among $x_{1}, \ldots, x_{n}, x_{n+1}$ and for
all $a_{1}, \ldots, a_{n}$ in $U$, 
there exists an element $b$ in $U$ such that for 
every $a_{n+1}$ in $U$,
$a_{n+1} \in_{\cal M} \sigma b$ if and only if 
$\llbracket A \rrbracket^{\cal U}_{a_{1}/x_{1}, \ldots, a_{n}/x_{n}, a_{n+1}/x_{n+1}} = 1$
\end{proposition}

\proof{
Let $|A|$ be the proposition defined as follows.
\begin{itemize}
\item $|A| = A$ if $A$ is atomic, 
\item $|A \Rightarrow B| = |A| \Rightarrow |B|$,
      $|A \wedge B| = |A| \wedge |B|$,
      $|A \vee B| = |A| \vee |B|$,
      $|\bot| = \bot$,
\item $|\forall x~A| = \forall x~((x \in E_{S(x)}) \Rightarrow |A|)$,
\item $|\exists x~A| = \exists x~((x \in E_{S(x)}) \wedge |A|)$.
\end{itemize}
Notice that the free variables of $|A|$ are among $E_{0}, \ldots, E_{m},
x_{1}, \ldots, x_{n}, x_{n+1}$. Let 
$$\varphi = a_{1} / x_{1}, \ldots, a_{n} / x_{n}, a_{n+1} / x_{n+1}$$
$$\psi = v_{0} / E_{0}, \ldots, v_{m} / E_{m}, \sigma^{k_{1}} a_{1} /
x_{1}, \ldots, \sigma^{k_{n}} a_{n} / x_{n}, \sigma^{k_{n+1}} a_{n+1} /
x_{n+1}$$ 
where $k_{1} = S(x_{1}), \ldots, k_{n+1} = S(x_{n+1})$. 
We check, by induction over the structure of $A$, that if $A$ is a
stratifiable proposition in the language $\in$, then 
$$\llbracket |A| \rrbracket^{\cal M}_{\psi} = \llbracket A \rrbracket^{\cal U}_{\varphi}$$

\begin{itemize}
\item If $A$ is an atomic proposition $x_{i} \in x_{j}$, then
$k_{j} = k_{i} + 1$, $\llbracket |A| \rrbracket^{\cal M}_{\psi} = 1$ if and only
if $\sigma^{k_{i}} a_{i} \in_{\cal M} \sigma^{k_{j}} a_{j}$ 
if and only if $a_{i} \in_{\cal M} \sigma a_{j}$, 
if and only if 
$\llbracket A \rrbracket^{\cal U}_{\varphi} = 1$.

\item if $A = B \Rightarrow C$ then 
$\llbracket |A| \rrbracket^{\cal M}_{\psi} = 1$ if and only if
$\llbracket |B| \rrbracket^{\cal M}_{\psi} = 0$ or
$\llbracket |C| \rrbracket^{\cal M}_{\psi} = 1$ 
if and only if $\llbracket B \rrbracket^{\cal U}_{\varphi} = 0$
or
$\llbracket C \rrbracket^{\cal U}_{\varphi} = 1$
if and only if 
$\llbracket A \rrbracket^{\cal U}_{\varphi} = 1$.

\item if $A = B \wedge C$ then 
$\llbracket |A| \rrbracket^{\cal M}_{\psi} = 1$
if and only if 
$\llbracket |B| \rrbracket^{\cal M}_{\psi} = 1$ 
and 
$\llbracket |C| \rrbracket^{\cal M}_{\psi} = 1$ if and only if
$\llbracket B \rrbracket^{\cal U}_{\varphi} = 1$ and
$\llbracket C \rrbracket^{\cal U}_{\varphi} = 1$ if and only if $\llbracket A \rrbracket^{\cal U}_{\varphi} = 1$.

\item if $A = B \vee C$ then 
$\llbracket |A| \rrbracket^{\cal M}_{\psi} = 1$
if and only if 
$\llbracket |B| \rrbracket^{\cal M}_{\psi} = 1$ 
or 
$\llbracket |C| \rrbracket^{\cal M}_{\psi} = 1$ if and only if
$\llbracket B \rrbracket^{\cal U}_{\varphi} = 1$ and
$\llbracket C \rrbracket^{\cal U}_{\varphi} = 1$ if and only if $\llbracket A \rrbracket^{\cal U}_{\varphi} = 1$.

\item $\llbracket |\bot| \rrbracket^{\cal M}_{\psi} = 0 = \llbracket \bot \rrbracket^{\cal U}_{\varphi}$.

\item if $A = \forall x~B$ then $\llbracket |A| \rrbracket^{\cal M}_{\psi} = 1$ 
if and only if for every $c$ in $M$ such that 
$c \in_{\cal M} v_{k}$, $\llbracket |B| \rrbracket^{\cal M}_{\psi + c/x} = 1$,
if and only if for every $e$ in $U$, 
$\llbracket |B| \rrbracket^{\cal M}_{\psi + \sigma^{k} e /x} = 1$
if and only if 
for every $e$ in $U$, 
$\llbracket B \rrbracket^{\cal U}_{\varphi + e/x} = 1$
if and only if 
$\llbracket A \rrbracket^{\cal U}_{\varphi} = 1$.

\item if $A = \exists x~B$ then $\llbracket |A| \rrbracket^{\cal M}_{\psi} = 1$ 
if and only if there exists $c$ in $M$ such that 
$c \in_{\cal M} v_{k}$
and $\llbracket |B| \rrbracket^{\cal M}_{\psi + c/x} = 1$,
if and only if there exists $e$ in $U$ such that 
$\llbracket |B| \rrbracket^{\cal M}_{\psi + \sigma^{k} e /x} = 1$
if and only if 
there exists $e$ in $U$ such that
$\llbracket B \rrbracket^{\cal U}_{\varphi + e/x} = 1$
if and only if 
$\llbracket A \rrbracket^{\cal U}_{\varphi} = 1$.
\end{itemize}

Then, the model ${\cal M}$ is a model of the comprehension
scheme. Hence, it is a model of the proposition 
$$\forall E_{0}~ \ldots~\forall E_{m}~\forall x_{1}~\ldots~\forall x_{n}~\forall y~\exists z~\forall x_{n+1}~
(x_{n+1} \in z \Leftrightarrow (x_{n+1} \in y \wedge |A|))$$
Thus, for all $a_{1}, ..., a_{n}$, 
there exists an object $b_{0}$ such that for all $a_{n+1}$
$$\llbracket (x_{n+1} \in z \Leftrightarrow (x_{n+1} \in y \wedge |A|))
\rrbracket^{\cal M}_{\psi + v_{k_{n+1}} / y + b_{0} / z} = 1$$
We have 
$\sigma^{k_{n+1}} a_{n+1} \in_{\cal M} b_{0}$
if and only if 
$\sigma^{k_{n+1}} a_{n+1} \in_{\cal M} v_{k_{n+1}}$ and $\llbracket |A|
\rrbracket^{\cal M}_{\psi} = 1$
thus 
$a_{n+1} \in_{\cal M} \sigma^{-k_{n+1}} b_{0}$
if and only if 
$a_{n+1}$ is in $U$ and $\llbracket A \rrbracket^{\cal U}_{\varphi} = 1$.
We take $b = \sigma^{-(k_{n+1}+1)} b_{0}$. For all $a_{n+1}$ in $U$, 
we have $a_{n+1} \in_{\cal M} \sigma b$
if and only if 
$\llbracket A \rrbracket^{\cal U}_{\varphi} = 1$.

Notice finally that $b_{0} \in_{\cal M} \wp_{\cal M} (v_{k_{n+1}})$, thus
$b_{0} \in_{\cal M} v_{k_{n+1}+1}$, $b \in_{\cal M} v_{0}$ and hence
$b$ is in $U$.}

\begin{definition}[Jensen's model]
The model ${\cal U} = \langle U, \in_{\cal U}, \hat{f}_{x_{1}, \ldots, x_{n},
y, A} \rangle$ 
is defined as follows. The base set is $U$. 
The relation $\in_{\cal U}$ is defined above.
The function 
$\hat{f}_{x_{1}, \ldots, x_{n}, x_{n+1}, A}$ maps $(a_{1}, \ldots, a_{n})$
to an object $b$ such that for all $a_{n+1}$ in $U$, 
$a_{n+1} \in_{\cal M} \sigma b$ if and only if $\llbracket A \rrbracket^{\cal U}_{a_{1}/x_{1}, \ldots, 
a_{n}/x_{n}, a_{n+1}/x_{n+1}} = 1$. 
\end{definition}

\begin{proposition}
The model ${\cal U}$ is a model of the Stratified Foundations. 
\end{proposition}

\proof{
If $A$ is a stratifiable proposition in the language $\in$, then
$$\llbracket 
t_{n+1} \in f_{x_{1}, \ldots, x_{n}, x_{n+1}, A}(t_{1}, \ldots, t_{n})
\rrbracket^{\cal U}_{\varphi} = 1$$ 
if and only if
$$\llbracket t_{n+1} \rrbracket^{\cal U}_{\varphi} \in_{\cal M} \sigma 
\hat{f}_{x_{1}, \ldots, x_{n}, x_{n+1}, A}
(\llbracket t_{1} \rrbracket^{\cal U}_{\varphi}, \ldots, \llbracket t_{n}
\rrbracket^{\cal U}_{\varphi})$$ 
if and only if 
$$\llbracket [t_{1}/x_{1}, \ldots, t_{n}/x_{n}, t_{n+1}/x_{n+1}]A
\rrbracket^{\cal U}_{\varphi} = 1$$ 
Hence, if $A \equiv B$ then  $A$ and $B$ have the same denotation.}

\begin{corollary}
The Stratified Foundations are consistent.
\end{corollary}

\subsection{Normalization}

We want now to construct a pre-model for the Stratified Foundations. 

Let $u_{i} = v_{3i}$ and $\tau = \sigma^{3}$. 
The function $\tau$ is an automorphism of ${\cal M}$, $\tau u_{i} =
u_{i+1}$, 
$u_{i} \subseteq_{\cal M} u_{i+1}$
and 
$\wp_{\cal M}~(\wp_{\cal M}~(\wp_{\cal M}~(u_{i}))) 
\subseteq_{\cal M} u_{i+1}$. 

As ${\cal M}$ is an $\omega$-model of set theory, for each recursively
enumerable relation $R$ on natural numbers, there is an object $r$ in
${\cal M}$ such that $R(a_{1}, \ldots, a_{n})$ if and only if
$\langle a_{1}, \ldots, a_{n} \rangle_{\cal M} \in_{\cal M} r$. 
In particular there is 
\begin{itemize}
\item 
an object $Proof$ such that $\pi \in_{\cal M} Proof$ if and only
if $\pi$ is (the encoding in ${\cal M}$ of the G\"{o}del number of) a
proof,
\item 
an object $Term$ such that $t \in_{\cal M} Term$ if and only
if $t$ is (the encoding of the G\"{o}del number of) a term,  
\item 
an object $Subst$ such that 
$\langle \pi, \pi_{1}, \alpha, \pi_{2} \rangle_{\cal M} \in_{\cal M} Subst$
if and only if $\pi, \pi_{1}$ and $\pi_{2}$ are 
(encodings of G\"{o}del numbers of) 
proofs, $\alpha$ is 
(the encoding of the G\"odel number of) 
a proof variable and $\pi = [\pi_{1}/\alpha]\pi_{2}$,
\item 
an object $Subst'$ such that 
$\langle \pi, t, x, \pi_{1} \rangle_{\cal M} \in_{\cal M} Subst'$
if and only if $\pi$ and $\pi_{1}$ are 
(encodings of the G\"{o}del numbers of) 
proofs, $x$ is 
(the encoding of the G\"odel number of) 
a term variable and $t$ 
(the encoding of the G\"odel number of) 
a term
and $\pi = [t/x]\pi_{1}$,
\item 
an object $Red$ such that $\langle \pi, \pi_{1} \rangle_{\cal M}
\in_{\cal M} Red$ if and only if $\pi$ and $\pi_{1}$ are 
(encodings of G\"odel numbers of) 
proofs and $\pi \triangleright^{*} \pi_{1}$,
\item 
an object $Sn$ such that $\pi \in_{\cal M} Sn$ if and only if
$\pi$ is 
(the encoding of the G\"odel number of) 
a strongly
normalizable proof, 
\item 
an object $ImpI$ such that $\langle \pi, \alpha, \pi_{1}
\rangle_{\cal M} \in_{\cal M} ImpI$ if and only if $\pi$  and
$\pi_{1}$ are 
(encodings of G\"odel numbers of) 
proofs, $\alpha$ is 
(the encoding of the G\"odel number of) 
a proof variable
and $\pi = \lambda \alpha~\pi_{1}$,
\item 
an object $AndI$ such that $\langle \pi, \pi_{1}, \pi_{2}
\rangle_{\cal M} \in_{\cal M} AndI$ if and only if $\pi, \pi_{1}$  and
$\pi_{2}$ are 
(encodings of G\"odel numbers of) 
proofs and $\pi = \langle \pi_{1}, \pi_{2} \rangle$,
\item 
an object $OrI1$ (resp. $OrI2$) such that 
$\langle \pi, \pi_{1} \rangle_{\cal M} \in_{\cal M} OrI1$ 
(resp.
$\langle \pi, \pi_{2} \rangle_{\cal M} \in_{\cal M} OrI2$) if and only if
$\pi$ and $\pi_{1}$ (resp. $\pi$ and $\pi_{2}$) are 
(encodings of G\"odel numbers of) 
proofs and  $\pi = i(\pi_{1})$ (resp. $\pi = j(\pi_{2})$),  
\item 
an object $ForallI$ such that $\langle \pi, \alpha, \pi_{1}
\rangle_{\cal M} \in_{\cal M} ForallI$ if and only $\pi$ and $\pi_{1}$
are 
(encodings of G\"odel numbers of) 
proofs, 
$\alpha$ is 
(the encoding of the G\"odel number of) 
a proof variable, 
and $\pi = \lambda \alpha \pi_{1}$, 
\item 
an object $ExistsI$ such that $\langle \pi, t, \pi_{1}
\rangle_{\cal M} \in_{\cal M} ExistsI$ if and only if 
$\pi$ and $\pi_{1}$
are 
(encodings of G\"odel numbers of) 
proofs, $t$ is (the encoding of the G\"odel number of) a term and $\pi
= \langle 
t, \pi_{1} \rangle$.
\end{itemize}

Notice also that, since ${\cal M}$ is a model of the comprehension
scheme, there is an object $Cr$ such that $\alpha \in_{\cal M} Cr$
if and only if $\alpha$ is a reducibility candidate (i.e. the set of
objects $\beta$ such that $\beta \in_{\cal M} \alpha$ is a
reducibility candidate). 

\begin{definition}[Admissible]
An element $\alpha$ of $M$ is said to {\em admissible} at level $i$ if
$\alpha$ is a set of pairs $\langle \pi,\beta
\rangle_{\cal M}$ where $\pi$ is a proof and $\beta$ an element of
$u_{i}$ and for each $\beta$ in $u_{i}$ the set of
$\pi$ such that $\langle \pi,\beta \rangle_{\cal M} \in_{\cal M} \alpha$ is a
reducibility candidate.
\end{definition}

Notice that if $R$ is any reducibility candidate then the set 
$R \times_{\cal M} u_{i}$ 
is admissible at level $i$. Hence there are admissible elements 
at all levels.

\begin{proposition}
There is an element $A_{i}$ in $M$ such that $\alpha
\in_{\cal M} A_{i}$ if and only if $\alpha$ is admissible at level $i$. 
\end{proposition}

\proof{An element $\alpha$ of ${\cal M}$ admissible at level $i$ if
and only if

\medskip
\noindent
$\alpha \in_{\cal M}
\wp_{\cal M}(Proof \times_{\cal M} u_{i})$

\hfill 
$\wedge \forall \beta~
(\beta \in_{\cal M} u_{i}
\Rightarrow \exists C~(C \in_{\cal M} Cr \wedge 
(\langle \pi,\beta \rangle_{\cal M} \in_{\cal M} \alpha
\Leftrightarrow \pi \in_{\cal M} C)))$

\medskip
\noindent
Hence, as ${\cal M}$ is a model of the comprehension scheme, there
is an element $A_{i}$ in $M$ such that $\alpha
\in_{\cal M} A_{i}$ if and only if $\alpha$ is admissible at level $i$.}

Notice that $\alpha \in \tau A_{i}$ if and only if 
$\alpha \in A_{i+1}$. Hence as ${\cal M}$ is a model of the
extensionality axiom, $\tau A_{i} = A_{i+1}$.  

Notice, at last, that 
$A_{i} 
\subseteq_{\cal M} 
\wp_{\cal M}(Proof \times_{\cal M} u_{i}) 
\subseteq_{\cal M} 
\wp_{\cal M}(u_{i} \times_{\cal M} u_{i}) 
\subseteq_{\cal M} 
\wp_{\cal M}~(\wp_{\cal M}~(\wp_{\cal M}~(u_{i}))) 
\subseteq_{\cal M} 
u_{i+1}$. 

\begin{proposition}
\label{candidat}
If $\beta \in_{\cal M} A_{i}$ and $\alpha \in_{\cal M} A_{i+1}$ then 
the set of $\pi$ such that $\langle\pi,\beta\rangle \in_{\cal M} \alpha$ is a
reducibility candidate. 
\end{proposition}

\proof{
As $\alpha \in_{\cal M} A_{i+1}$ and $\beta \in_{\cal M} A_{i}
\subseteq_{\cal M} u_{i+1}$, 
the set of $\pi$ such that $\langle\pi,\beta\rangle
\in_{\cal M} \alpha$ is a reducibility candidate.} 

\medskip

We are now ready to construct a pre-model ${\cal N}$ of the Stratified
Foundations. The base set of this pre-model is the set $N$ of elements
of $M$ that are admissible at level $0$. We take 
$\in_{\cal N}(\alpha,\beta) = 
\{\pi~|~\langle \pi,\alpha \rangle_{\cal M} \in_{\cal M} \tau \beta\}$.  
This permits to define the
denotation of propositions built without Skolem symbols.  To define the
denotation of Skolem symbols, we prove the following proposition.

\begin{proposition}
For every stratifiable proposition $A$ in the language $\in$
whose free variables are among $x_{1}, \ldots, x_{n}, x_{n+1}$ and for
all $a_{1}, \ldots, a_{n}$ in $N$, 
there exists an element $b$ in $N$ such that for 
every $a_{n+1}$ in $N$,
$\langle \pi, a_{n+1}
\rangle_{\cal M} \in_{\cal M} \tau b$ if and only if 
$\pi$ is in $\llbracket A \rrbracket^{\cal N}_{a_{1}/x_{1}, \ldots, a_{n+1}/x_{n+1}}$.
\end{proposition}

\proof{
Let $|A|$ be the proposition (read {\em $p$ realizes $A$})
defined as follows.
\begin{itemize}
\item $|x_{i} \in x_{j}| = \langle p,x_{i} \rangle \in x_{j}$,

\item $|A \Rightarrow B| = p \in sn \wedge
\forall q~\forall w~\forall r~(\langle p, q \rangle \in red
\wedge \langle q,w,r \rangle \in impI) \Rightarrow \forall s~
[s/p]|A| \Rightarrow \forall t~\langle t,s,w,r \rangle \in subst
\Rightarrow [t/p]|B|)$, 

\item $|A \wedge B| = p \in sn \wedge
\forall q~\forall r~\forall s~(
(\langle p,q \rangle \in red \wedge \langle q,r,s \rangle \in andI)
\Rightarrow [r/p]|A| \wedge [s/p]|B|)$,  

\item $|A \vee B| = p \in sn \wedge
\forall q~\forall r~(
(\langle p,q \rangle \in red \wedge \langle q,r \rangle \in orI1)
\Rightarrow [r/p]|A|)
\wedge
\forall q~\forall r~(
(\langle p,q \rangle \in red \wedge \langle q,r \rangle \in orI2)
\Rightarrow [r/p]|B|)$,  

\item $|\bot| = p \in sn$,

\item $|\forall x~A| = p \in sn \wedge 
\forall q~\forall w~\forall r~(\langle p,q \rangle \in red \wedge (
\langle q,w,r \rangle \in forallI)
\Rightarrow \forall x~\forall y~(x \in E_{S(x)} \wedge y \in term)
\Rightarrow \forall s~(\langle s,w,y,r \rangle \in subst' \Rightarrow
[r/p,x/x]|A|))$,  

\item $|\exists x~A| = p \in sn \wedge 
\forall q~\forall t~\forall r~(\langle p,q \rangle \in red \wedge (
\langle q,t,r \rangle \in existsI)
\Rightarrow \exists x~x \in E_{S(x)}
\Rightarrow [r/p,x/x]|A|))$.  

\end{itemize}

Notice that the free variables of $|A|$ are among 
$term, subst, subst', red, sn,$ $impI, andI, orI1, orI2, forallI,
existsI,$ $p, E_{0}, \ldots, E_{m},
x_{1}, \ldots, x_{n}, x_{n+1}$. Let 
$$\varphi = a_{1} / x_{1}, \ldots, a_{n} / x_{n}, a_{n+1} / x_{n+1}$$
$$\psi = 
Term/term, Subst/subst, Subst'/subst', Red/red, Sn/sn,$$
$$ImpI/impI, AndI/andI, OrI1/orI1, OrI2/orI2, ForallI/forallI,
ExistsI/existsI,$$
$$A_{0} / E_{0}, \ldots, A_{m} / E_{m}, \tau^{k_{1}} a_{1} /
x_{1}, \ldots, \tau^{k_{n}} a_{n} / x_{n}, \tau^{k_{n+1}} a_{n+1} /
x_{n+1}$$ 

We check, by induction over the structure of $A$, that if $A$ is a
stratifiable proposition in the language $\in$, then the set of proofs $\pi$ such that  
$\llbracket |A| \rrbracket^{\cal M}_{\psi + \pi/p} = 1$ is 
$\llbracket A \rrbracket^{\cal N}_{\varphi}$.

\begin{itemize}
\item If $A$ is an atomic proposition $x_{i} \in x_{j}$, then $k_{j} =
k_{i} + 1$, 
we have $\llbracket |A| \rrbracket^{\cal M}_{\psi+\pi/p} = 1$ 
if and only if $\langle \pi, \tau^{k_{i}} a_{i} \rangle_{\cal M}
\in_{\cal M} \tau^{k_{j}} a_{j}$ 
if and only if $\langle \tau^{k_{i}} \pi, \tau^{k_{i}} a_{i} \rangle_{\cal M}
\in_{\cal M} \tau^{k_{j}} a_{j}$ 
if and only if 
$\tau^{k_{i}} \langle \pi, a_{i} \rangle_{\cal M} \in_{\cal M}
\tau^{k_{j}} a_{j}$  
if and only if 
$\langle \pi, a_{i} \rangle_{\cal M} \in_{\cal M} \tau a_{j}$ 
if and only if 
$\pi$ is in $\llbracket A \rrbracket^{\cal N}_{\varphi}$.

\item if $A = B \Rightarrow C$ then we have $\llbracket |A|
\rrbracket^{\cal M}_{\psi+\pi/p} = 1$ if and only if $\pi$ is strongly
normalizable and whenever $\pi$ reduces to $\lambda \alpha~\pi_{1}$ then
for all $\pi'$ such that $\llbracket |B| \rrbracket^{\cal M}_{\psi+
\pi'/p} = 1$ we have $\llbracket |C| \rrbracket^{\cal M}_{\psi +
[\pi'/\alpha]\pi_{1}/p} = 1$ if and only if $\pi$ is strongly
normalizable and whenever $\pi$ reduces to $\lambda x~\pi_{1}$ then
for all $\pi'$ in $\llbracket B \rrbracket^{\cal N}_{\varphi}$, $[\pi'/\alpha]\pi_{1}$ is in
$\llbracket C \rrbracket^{\cal N}_{\varphi}$ if and only if $\pi$ is in $\llbracket A \rrbracket^{\cal N}_{\varphi}$.

\item
If $A = B \wedge C$ then we have $\llbracket A \rrbracket^{\cal M}_{\psi+\pi/p}
= 1$ if and only if $\pi$ is strongly normalizable and whenever $\pi$
reduces to $\langle \pi_{1},\pi_{2} \rangle$ then $\llbracket
B\rrbracket^{\cal M}_{\psi+\pi_{1}/p} = 1$ and $\llbracket C
\rrbracket^{\cal M}_{\psi+\pi_{2}/p} = 1$ if and only if $\pi$ is strongly
normalizable and whenever $\pi$ reduces to $\langle \pi_{1},\pi_{2}
\rangle$ then $\pi_{1}$ is in $\llbracket B \rrbracket^{\cal N}_{\varphi}$ and $\pi_{2}$ is in
$\llbracket C \rrbracket^{\cal N}_{\varphi}$
if and only if $\pi$ is in $\llbracket A \rrbracket^{\cal N}_{\varphi}$.

\item
If $A = B \vee C$ then we have
$\llbracket A \rrbracket^{\cal M}_{\psi+\pi/p} = 1$ 
if and only if 
$\pi$ is strongly normalizable and whenever
$\pi$ reduces to $i(\pi_{1})$ (resp. $j(\pi_{2})$) then
$\llbracket B \rrbracket^{\cal M}_{\psi+\pi_{1}/p} = 1$ 
(resp. $\llbracket C \rrbracket^{\cal M}_{\psi+\pi_{2}/p} = 1$)
if and only if 
$\pi$ is strongly normalizable and whenever
$\pi$ reduces to $i(\pi_{1})$ (resp. $j(\pi_{2})$) then $\pi_{1}$ is
in $\llbracket B \rrbracket^{\cal N}_{\varphi}$ (resp. $\llbracket C \rrbracket^{\cal N}_{\varphi}$)
if and only if $\pi$ is in $\llbracket A \rrbracket^{\cal N}_{\varphi}$.

\item
If $A = \bot$ then 
$\llbracket A\rrbracket^{\cal M}_{\psi + \pi/p} = 1$ if and only if 
$\pi$ is strongly normalizable 
if and only if $\pi$ is in $\llbracket A \rrbracket^{\cal N}_{\varphi}$. 

\item if $A = \forall x~B$, then 
$\llbracket |A| \rrbracket^{\cal M}_{\psi + \pi/p} = 1$ if
and only if  
$\pi$ is strongly normalizable and whenever $\pi$ reduces to 
$\lambda x~\pi_{1}$, for all term $t$ and for all $c$ in $M$ such that 
$c \in_{\cal M} A_{k}$, 
$\llbracket |B| \rrbracket^{\cal M}_{\psi +  c / x, [t/x]\pi_{1}/p} = 1$
if and only if 
$\pi$ is strongly normalizable and whenever $\pi$ reduces to 
$\lambda x~\pi_{1}$, for all $t$ and for all $e$ in $N$, 
$\llbracket |B| \rrbracket^{\cal M}_{\psi + \tau^{k} e / x +
[t/x]\pi_{1}/p} = 1$ 
if and only if 
$\pi$ is strongly normalizable and whenever $\pi$ reduces to 
$\lambda x~\pi_{1}$, for all $t$ and for all $e$ in $N$, 
$[t/x]\pi_{1}$ is in $\llbracket B \rrbracket^{\cal N}_{\varphi + e/x}$
if and only if 
$\pi$ is in $\llbracket A \rrbracket^{\cal N}_{\varphi}$.

\item if $A = \exists x~B$, then 
$\llbracket |A| \rrbracket^{\cal M}_{\psi + \pi/p} = 1$ if
and only if  
$\pi$ is strongly normalizable and whenever $\pi$ reduces to 
$\langle t,\pi_{1} \rangle$, there exists a $c$ in $M$ such that 
$c \in_{\cal M} A_{k}$ and 
$\llbracket |B| \rrbracket^{\cal M}_{\psi +  c / x, [t/x]\pi_{1}/p} = 1$
if and only if 
$\pi$ is strongly normalizable and whenever $\pi$ reduces to 
$\langle t,\pi_{1} \rangle$, there exists a $e$ in $N$ such that 
$\llbracket |B| \rrbracket^{\cal M}_{\psi + \tau^{k} e / x +
[t/x]\pi_{1}/p} = 1$ 
if and only if 
$\pi$ is strongly normalizable and whenever $\pi$ reduces to 
$\langle t,\pi_{1} \rangle$, there exists a $e$ in $N$ such that
$[t/x]\pi_{1}$ is in $\llbracket B \rrbracket^{\cal N}_{\varphi + e/x}$
if and only if 
$\pi$ is in $\llbracket A \rrbracket^{\cal N}_{\varphi}$.
\end{itemize}

Then, the model ${\cal M}$ is a model of the comprehension
scheme. Hence, it is a model of the proposition 
$$
\forall E_{0}~ \ldots~\forall E_{m}~\forall x_{1}~\ldots~\forall x_{n}~\exists z~\forall
p~\forall x_{n+1}~ 
\langle p, x_{n+1} \rangle \in z \Leftrightarrow 
\langle p, x_{n+1} \rangle \in proof \times U \wedge |A|$$
Thus, for all $a_{1}, ..., a_{n}$, 
there exists an object $b_{0}$ such that for all $a_{n+1}$
$$\llbracket 
\langle p, x_{n+1} \rangle \in z \Leftrightarrow 
\langle p, x_{n+1} \rangle \in {\mathbb N}_{\cal M} \times U \wedge |A|
\rrbracket^{\cal M}_{\psi + Proof/proof, b_{0}/z, u_{k_{n+1}+1} / U, \pi/p} = 1$$

We have 
$\langle \pi, \tau^{k_{n+1}} a_{n+1} \rangle_{\cal M} \in_{\cal M}
b_{0}$ 
if and only if $\pi$ is a proof,
$\tau^{k_{n+1}} a_{n+1} \in_{\cal M} u_{k_{n+1}+1}$ and 
$\llbracket |A| \rrbracket^{\cal M}_{\psi + \pi/p} = 1$. 
Thus 
$\langle \pi, a_{n+1} \rangle_{\cal M} \in_{\cal M}
\tau^{-k_{n+1}} b_{0}$ 
if and only if 
$a_{n+1} \in_{\cal M} u_{1}$ and $\pi$ is in $\llbracket A \rrbracket^{\cal N}_{\varphi}$. 
We take $b = \tau^{-(k_{n+1} + 1)} b_{0}$ and for all $a_{n+1}$ in $N$
we have 
$\langle \pi, a_{n+1} \rangle_{\cal M} \in_{\cal M}
\tau b$ 
if and only if $\pi$ is in $\llbracket A \rrbracket^{\cal N}_{\varphi}$. 
Finally, notice that 
$b_{0}$ is a set of pairs $\langle \pi,\beta \rangle_{\cal M}$ where
$\pi$ is a proof and $\beta$ an element of $u_{k_{n+1}+1}$
and for each $\beta$ in $u_{k_{n+1}+1}$ the set of
$\pi$ such that $\langle \pi,\beta \rangle_{\cal M} \in_{\cal M}
b_{0}$ is 
$\llbracket |A| \rrbracket^{\cal M}_{\psi + \beta/x_{k_{n+1}}, \pi/p} = 1$,
hence it is a 
reducibility candidate. 
Hence $b_{0} \in_{\cal M} A_{k_{n+1}+1}$ and $b$ is in $N$.}

\begin{definition}[Crabb\'{e}'s pre-model]

The pre-model ${\cal N} = \langle N, \in_{\cal N}, \hat{f}_{x_{1}, \ldots,
x_{n}, y, A} \rangle$ is defined as follows. The base set is $N$.  
The function $\in_{\cal N}$ is defined above. 
The function 
$\hat{f}_{x_{1}, \ldots, x_{n}, x_{n+1}, A}$ maps $(a_{1}, \ldots, a_{n})$
to the object $b$ such that for all $a_{n+1}$ in $N$, 
$\langle \pi, a_{n+1} \rangle_{\cal M} \in_{\cal M} \tau b$ if and only if 
$\pi$ is in $\llbracket A \rrbracket^{\cal N}_{a_{1}/x_{1}, \ldots, a_{n}/x_{n}, a_{n+1}/x_{n+1}}$.
\end{definition}

\begin{proposition}
The pre-model ${\cal N}$ is a pre-model of the Stratified Foundations. 
\end{proposition}

\proof{
If $A$ is a stratifiable proposition 
in the language $\in$, then 
$$\mbox{$\pi$ is in $\llbracket 
t_{n+1} \in f_{x_{1}, \ldots, x_{n}, x_{n+1}, A}(t_{1}, \ldots, t_{n})
\rrbracket^{\cal N}_{\varphi}$}$$
if and only if
$$\langle \pi, \llbracket t_{n+1} \rrbracket^{\cal N}_{\varphi} \rangle_{\cal
M} \in_{\cal M} \tau \hat{f}_{x_{1}, \ldots, x_{n}, x_{n+1}, A}
(\llbracket t_{1} \rrbracket^{\cal N}_{\varphi}, \ldots, \llbracket t_{n}
\rrbracket^{\cal N}_{\varphi})$$
if and only if 
$$\mbox{$\pi$ is in $\llbracket [t_{1}/x_{1}, \ldots, t_{n}/x_{n},
t_{n+1}/x_{n+1}]A \rrbracket^{\cal N}_{\varphi}$}$$
Hence, if $A \equiv B$ then  $A$ and $B$ have the same denotation.}

\begin{corollary}
All proofs strongly normalize in the Stratified Foundations.
\end{corollary}

\begin{remark}
As already noticed in \cite{Crabbe91}, instead of constructing the a
pre-model of the Stratified Foundations within an automorphic
$\omega$-model of Zermelo's set theory, we could construct it within
an $\omega$-model of the Stratified Foundations.
In such a model ${\cal U}$, we can define recursively enumerable
relations, because the Stratified Foundations contains enough
arithmetic and comprehension. Then we can take the sequence $u_{i}$ to
be the constant sequence equal to $w$ where $w$ is a universal set,
i.e. a set such that $a \in_{\cal U} w$ for all element $a$ of the
model. Such an object obviously verifies 
$\wp_{\cal U}(\wp_{\cal U}(\wp_{\cal U}(w))) \subseteq_{\cal U} w$.
In other words, we say that an element
of $U$ is admissible if it is a set of pairs $\langle \pi, \beta
\rangle_{\cal U}$ where $\pi$ is a proof and for each $\beta$ in $U$,
the set of $\pi$ such that $\langle \pi, \beta \rangle \in_{\cal U}
\alpha$ is a reducibility candidate. Proposition \ref{candidat}
becomes trivial, but we need to use the existence of a universal set
to prove that there are admissible elements in the model and that there is 
a set $A$ of admissible elements in the model.
Hence, the difficult part in this pre-model construction (the part
that would not go through for Zermelo's set theory for instance) is
the construction of the base set.
\end{remark}

\section*{Conclusion}

In this paper, we have have shown that the Stratified Foundations can
be expressed in deduction modulo and that the normalization proof for
this theory be decomposed into two lemmas: one expressing that it has
a pre-model and the other that proof normalize in this theory if it
has a pre-model. This second lemma is not specific to the Stratified
Foundations, but holds for all theories modulo.
The idea of the first lemma is to construct a pre-model within an
$\omega$-model of the theory with the help of formal
realizability. 
This idea does not seems to be specific to the Stratified Foundations
either, but, its generality remains to be investigated.  Thus, this
example contributes to explore of the border between the theories
modulo that have the normalization property and those that do not.

\end{document}